\documentclass[twocolumn,showpacs,preprintnumbers,amsmath,amssymb,prl]{revtex4}
\usepackage{graphicx}% Include figure files
\usepackage{dcolumn}% Align table columns on decimal point
\usepackage{bm}% bold math
\usepackage{epsf,epsfig,latexsym}

\begin{document}
%\preprint{APS/123-QED}
\title{Limiting temperatures and the Equation of State of Nuclear Matter}

\author{J. B. Natowitz}
\author{K. Hagel}
\author{Y. Ma}
\author{M. Murray}
\author{L. Qin}
\author{R. Wada}
\author{J. Wang}
\affiliation{Cyclotron Institute, Texas A\&M University,\\
  College Station,  Texas, 77845}%

\date{\today}% It is always \today, today           %  but any date may be explicitly specified

\begin{abstract}
From experimental observations of limiting temperatures in heavy ion 
collisions we derive $T_c$, the critical temperature of infinite nuclear
 matter. The critical temperature is $16.6 \pm 0.86$ MeV. Theoretical model 
correlations between $T_c$, the compressibility modulus, $K$ the effective 
mass, $m^*$ and the saturation density, $\rho_s$, are exploited to derive 
the quantity $(K/m^*)^\frac{1}{2}\rho_s^{-\frac{1}{3}}$. This quantity 
together with calculations employing Skyrme and Gogny interactions 
indicates a nuclear matter incompressibility in moderately excited 
nuclei that is in excellent agreement with the value determined from 
Giant Monopole Resonance data.  This technique of extraction of $K$ may 
prove particularly useful in investigations of very neutron rich systems 
using radioactive beams.
\end{abstract}

\maketitle

Improved knowledge of the nuclear equation of state and a coherent picture 
of the 
relationship between the properties of finite nuclei and bulk nuclear 
matter remains a key 
requirement in both nuclear physics and astrophysics. It is key, for 
example, to  
understanding nuclear structure, heavy ion collisions, supernova explosions 
and neutron 
star properties\cite{bali98,lattimer00,reinhard00}. Significant effort has been devoted to the development of 
microscopic theoretical models which can provide reliable mathematical 
formulations of 
this equation of state\cite{blaizot80,jaqaman84,myers98,zimanyi90,%
reinhard86,blaizot95,furnstahl97,lalazissis97,malheiro96,chabanat97,%
farine97,furnstahl84,goodman84,boguta77,von-eiff94,onsi02,martins00}. 
Such calculations are usually specified 
for symmetric 
nuclear matter, a hypothetical system of equal numbers of neutrons and 
(uncharged) 
protons interacting through nuclear forces. Driven by the astrophysical 
problems and 
more recent laboratory excursions into the region of more exotic nuclei, 
the dependence 
of the equation of state on neutron-proton asymmetry has also become a 
subject of 
significant interest.\cite{muller95,chung01,haddad01}. In this letter we 
employ data from experimental measurements of caloric curves in nuclear 
collisions, together with systematic trends and 
correlations derived from a number of theoretical investigations of 
nuclear matter, to derive the critical temperature and incompressibility of 
symmetric nuclear matter. The techniques employed offer a natural method 
to extend such investigations to more asymmetric systems.

In a recent paper measurements of nuclear specific heats from a large 
number of 
experiments were employed to construct caloric curves for five different 
regions of nuclear mass\cite{natowitz02}. Within experimental 
uncertainties each of these caloric curves exhibits 
a plateau region at higher excitation energy, i.e., a ``limiting 
temperature'' is reached. In Figure 1 these limiting temperatures from 
reference\cite{natowitz02} are presented as a function of mass. As 
previously noted, they are observed to decrease with increasing 
mass. This decrease with increasing mass has long been predicted as 
resulting from Coulomb Instabilities of expanded and heated 
nuclei\cite{bonche85,levit85,besprovany89,song91,zhang96_1,zhang96_2,%
das92,song93,song94,baldo99,zhang99}. 

The results employed in reference\cite{natowitz02} were based upon 
temperature determinations derived from double isotope yield ratios and 
from slope measurements of particle spectra. More recently the TAPS 
Collaboration has reported temperatures determined 
from a new technique, observations of ``second chance'' bremsstrahlung 
gamma ray emission for a series of reactions which span a wide range of 
mass\cite{enterria02_1,enterria02_2}. There are not yet sufficient data 
of this latter type to construct caloric curves for relatively narrow mass 
regions as was done for the previous temperature data. However, in each 
case studied with this technique the collisions lead to excitation energies 
which are above those 
identified as the starting points of the plateau regions identified in 
reference\cite{natowitz02}. Thus it is 
reasonable to compare the temperatures determined from the thermal 
bremsstrahlung  
measurements with the earlier limiting temperature values. As seen in 
Figure 1, the 
reported second chance gamma temperatures and their mass dependence 
are in excellent 
agreement with the earlier results. We take this agreement as an 
independent 
confirmation of the earlier results and note that the new results 
extend the determination 
of the mass dependence to significantly higher mass.  

A relatively large number of theoretical calculations of the critical 
temperature of semi-
infinite nuclear matter (nuclear matter with a surface)  have been 
reported in the literature \cite{bonche85,levit85,%
besprovany89,song91,zhang96_1,zhang96_2,das92,song93,song94,baldo99,zhang99}.
%[25-35 and references therein]. 
The different nuclear interactions employed in these calculations lead to 
large differences in the critical temperatures derived. Values from 13 
to 24 MeV are reported in references \onlinecite{bonche85,levit85,%
besprovany89,song91,zhang96_1,zhang96_2,das92,song93,song94,baldo99,zhang99}. 
The limiting temperatures plotted in Figure 1 are well below these 
calculated critical temperatures. This difference reflects finite size 
effects, Coulomb effects and isospin asymmetry effects for the finite 
nuclei studied.  A first order estimate of the magnitude of these combined 
effects can be made by comparing the volume coefficient of the Liquid 
Drop Model Binding Energy Equation, -16.0 MeV, which represents the 
binding energy per nucleon in infinite nuclear matter, to typical nuclear 
binding energies, $\approx 8$ MeV/nucleon. This suggests that limiting 
temperatures in nuclei should be $\approx 0.5$ times the critical temperature 
of nuclear matter \cite{jaqaman84}. Given the wide variation in the 
calculated values of $T_c$ it is not surprising that large 
variations result for the absolute values of limiting temperatures 
calculated for finite nuclei.

Employing a variety of Skyrme type interactions Song and Su have 
previously noted a mass dependent scaling of the Coulomb Instability 
temperatures with the critical temperature of nuclear matter 
(see Figure 6 of reference \onlinecite{song91}). These calculations 
were performed for nuclei along the line of beta stability. A similar 
scaling exists when other model interactions are employed. 

Mean values of $T_{lim}/T_c$ for five different masses which result from 
averaging the results of different calculations \cite{bonche85,levit85,%
besprovany89,song91,zhang96_1,zhang96_2,das92,song93,song94,baldo99,zhang99}
are shown in Figure 2. The estimated uncertainties are relatively 
small, $\approx 6\%$. For comparison, the figure also presents ratios 
of $T_{lim}/T_c$ which are expected to result assuming only finite size 
effects as derived from a lattice calculation\cite{elliott00} and the 
ratio of the nuclear binding energy per nucleon along the line of 
beta stability to the bulk binding energy per nucleon, 16 MeV. We have 
employed the mean variation of $T_{lim}/T_c$ with A, determined from 
commonly used microscopic theoretical calculations, together with the 
five experimental limiting temperatures reported in 
reference \onlinecite{natowitz02}, to extract the critical temperature of 
nuclear matter. In doing so we treat the theoretical variation as if it were 
an experimental uncertainity. Since the various interactions employed 
have been ``tuned'' to other nuclear properties, we consider this a 
reasonable approach.  The results are presented in Figure 3. Averaging the 
individual results we find $16.6 \pm 0.86$ MeV.  

It is interesting to ask whether additional Equation of State 
information can be extracted from this result. Blaizot 
{\it et al.}\cite{blaizot95} have argued that the most effective way to 
extract the 
incompressibility modulus of nuclear matter from experimental data is 
by comparison with microscopic calculations. This is usually done by 
comparison of the measured energies for the centroids of the strength 
distribution of Giant Monopole Resonances (GMR) with the calculated 
centroids. The generally accepted best current value of $K = 231 \pm 5$ MeV 
has been determined in such a fashion\cite{youngblood99} by comparison 
with the calculated centroids using Gogny interactions\cite{blaizot95}. 
We have adopted a similar comparison procedure using the present result 
for $T_c$ determination.

We began by using a relation suggested by the work of Kapusta\cite{kapusta84}
and Lattimer and Swesty \cite{lattimer91} who have pointed out that 
correlations between parameters used to describe nuclear matter are such 
that a relationship between the critical temperature, $T_c$, the 
incompressibility, $K$, the effective mass, 
$m^*$ (= $m_{eff}/m$ where $m_{eff}$ is the nucleon effective mass and 
$m$ is the nucleon mass) and the saturation density, 
$\rho_s$, may be written as

\begin{eqnarray}
T_c = C_T  (K/m^*)^\frac{1}{2}\rho_s^{-\frac{1}{3}}
\end{eqnarray}

\noindent where $C_T$ is a constant. Using this relationship, we have 
determined the constant $C_T$ in this equation using published theoretical 
values for $T_c$ calculated utilizing a number of 
different microscopic interactions \cite{blaizot80,jaqaman84,myers98,%
zimanyi90,reinhard86,blaizot95,furnstahl97,lalazissis97,malheiro96,chabanat97,%
farine97,furnstahl84,goodman84,boguta77,von-eiff94,onsi02,martins00}. 
Results of calculations using interactions with $155 < K <384$ MeV 
are depicted in Figure 4.  A least squares fit to these data suggests 
a very slight decrease of $C_T$ with increasing 
$(K/m^*)^\frac{1}{2}\rho_s^{-\frac{1}{3}}$. Using the present value for $T_c$ 
and an iterative technique to establish $C_T$ leads to 
$$C_T =  0.484 \pm 0.074$$
\noindent and
$$(K/m^*)^\frac{1}{2}\rho_s^{-\frac{1}{3}}=34.2\pm 5.34 MeV^\frac{1}{2} fm.$$

\noindent The saturation density, $\rho_s$, is well established by 
charge density measurements to be $0.16 \pm 0.005$ fm$^{-3}$\cite{chabanat97}. 
The standard deviation of the model values of $\rho_s$ from 0.16 fm$^{-3}$, 
calculated with the different interactions, is 3.4\%. 
Either of these uncertainties is very small compared to other uncertainties 
in the determination.  Therefore $T_c$ is a measure of $K/m^*$. 
It is important to recognize that $K$ and $m^*$ are not independent 
variables but are correlated\cite{blaizot95,chabanat97}. 
For Skyrme effective interactions Chabanet {\it et al.} have 
given expressions for $K$ and $m^*$ and discussed the resultant correlation 
between them \cite{chabanat97}. Using the relationships discussed in 
that work $m^*$ can be written as:
\begin{eqnarray}
m^*=(1 + \frac{m}{8\hbar^2}\rho\Theta_s)^{-1}
\end{eqnarray}

\noindent where 
\begin{eqnarray}
\Theta_s = \frac{K-B-C\sigma}{D(1-\frac{3}{2}\sigma)}
\end{eqnarray}

and $\sigma$ is a parameter which ranges from 0 to 1 and controls the density 
dependence of the interaction.  B, C, D  are parameters directly related 
to $e_\infty$, the energy per nucleon in 
infinite nuclear matter and $e_F$ , the Fermi energy of infinite nuclear 
matter.
\begin{eqnarray}
B =  -9e_\infty + \frac{3}{5}e_F 
\end{eqnarray}
\begin{eqnarray}
C =  -9e_\infty + \frac{9}{5}e_F 
\end{eqnarray}
\begin{eqnarray}
D =  \frac{3}{20}\rho k_F^2    
\end{eqnarray}

\noindent Here $k_F$ is the Fermi momentum.

It is clear then that for a given K the ratio $K/m^*$ in equation 1  
depends on the choice of $\sigma$, the parameter of the density dependent 
term.  (In the Gogny interactions of reference \onlinecite{blaizot95} this 
parameter controlling the density dependent term is designated $\alpha$.). 
As a result, determination of $K$ from $K/m^*$ is sensitive to the choice 
of this parameter.  For example in reference \onlinecite{chabanat97} 
the relation between $K$ and $m^*$ is such that small values of $\sigma$ 
dictate lower values of $K$.  Also, for smaller values of $\sigma$, $m^*$ 
decreases as $K$ increases while for larger $\sigma$, $m^*$ increases 
with increasing K.(See Fig 2, Reference \onlinecite{chabanat97}).

For comparison to the data we present in Figure 5 a plot of $K$ vs 
$(K/m^*)^\frac{1}{2}\rho_s^{-\frac{1}{3}}$, obtained using the Gogny 
interactions from reference \onlinecite{blaizot95} and various Skyrme 
interactions. The dashed lines in the plot show the trend of the 
generalized Skyrme interactions for $\sigma = \frac{1}{6}, \frac{1}{3}$ 
and $1$, as obtained in reference \cite{chabanat97}. The solid lines 
connect results for Gogny interactions with $\alpha = \frac{1}{3}$ 
and $\frac{2}{3}$\cite{blaizot95}.  As seen in the 
figure, a higher value of $\sigma$ leads to a higher apparent $K$. 
It has been pointed out that maintaining $K$ in a ``reasonable'' range of 
200-300 MeV requires low values of $\sigma$ 
(or $\alpha$)\cite{jaqaman84,blaizot95,chabanat97}. In particular, 
the value of $K = 231 \pm 5$ MeV derived from the GMR data\cite{youngblood99}
was obtained by comparison of data for the breathing mode energy of five 
different nuclei with energies calculated employing the 
Gogny D1 ($\alpha = \frac{1}{3}$), D1S($\alpha = \frac{1}{3}$) and 
D250($\alpha = \frac{2}{3}$) interactions\cite{blaizot95}. 
For three of these nuclei only the D1S interaction results were used. 
For the other two a fit to the trend in energies calculated from the 
three interactions was employed. 

The value of $(K/m^*)^\frac{1}{2}\rho_s^{-\frac{1}{3}}$ derived from this 
work is also indicated on the Figure by the vertical line. We note that 
this line intersects the calculated values essentially at a point 
where the $\frac{1}{6}$ Skyrme and $\frac{1}{3}$ Gogny lines intersect. 
The different slopes of the Skyrme and Gogny lines in Figure lead to 
different uncertainties in the K value. Thus employing Skyrme interactions 
with the $\sigma = \frac{1}{6}$ parameterization\cite{chabanat97}, 
$K = 232 \pm 22$MeV. Using Gogny interactions with 
$\alpha = \frac{1}{3}$\cite{blaizot95} leads to $K = 233 \pm 39$ MeV. 
These results for $K$ lead respectively to $m^*$ values of 
%$0.674 \pm 0.06$ 
$0.674^{+.18}_{-.13}$
or $0.674^{+0.11}_{-0.09}$. The compressibility modulus 
determined from the critical temperature in this manner is then entirely 
consistent with that determined from the GMR measurements. Higher 
values of $\sigma$ (or $\alpha$) will lead to higher apparent $K$. 
Thus for the Skyrme $\sigma = \frac{1}{3}$ line a value of $K = 252$ 
would result. For the extension of the Gogny $\alpha = \frac{2}{3}$ line, 
$K = 242$ would be obtained).  The calculated breathing mode energies 
are apparently less sensitive to the value of the parameter of the 
density dependent interaction.

In summary, from limiting temperature values obtained in five different 
mass regions we have determined a critical temperature of $16.6 \pm 0.86$ MeV 
for symmetric infinite nuclear matter. This has been used to derive both $K$, 
the incompressibility and $m^*$, the effective mass. Extracted by comparison 
with the same interactions as were employed to determine K from observations 
of the Giant Monopole Resonance at low excitation energy, the 
value of K, obtained here from properties of nuclei at moderate 
excitation energies, is found to be in excellent agreement with that 
GMR result\cite{youngblood99}. The precision of the GMR measurement is 
better than that obtained from the present determination which 
incorporates data from a number of different experiments. The precision 
for the $T_c$ measurement could be improved. However, given the relative 
complexity of the collision dynamics involved, the breathing mode 
measurements should remain as the standard.  Nevertheless, using newly 
available radioactive beams the determination of limiting and  
critical temperatures may play a significant role in providing a means 
to establish the N/Z asymmetry dependence of the compressibility modulus 
and other important nuclear properties\cite{bali01}.

\section{Acknowledgements}

We appreciate helpful discussions with C. A. Gagliardi, J. Hardy, 
S. Shlomo and D. H. Youngblood. This work was supported by the 
U S Department of energy under Grant DE-FG03-93ER40773 and by the 
Robert A. Welch Foundation.

\begin{figure}
\epsfig{file=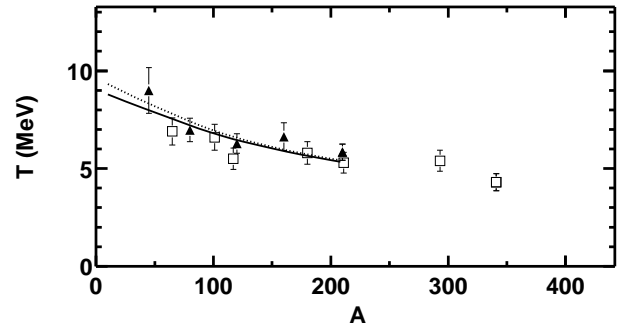,width=9.2cm,angle=0}
\caption{\label{fg:LimitingTemperatures}
Limiting Temperatures vs Mass. Limiting temperatures derived from double 
isotope yield ratio measurements are represented by solid triangles. 
Temperatures derived from thermal bremsstrahlung measurements are 
represented by open squares.  Lines represent limiting temperatures 
calculated using interactions proposed by Gogny (dashed)\cite{zhang96_2} 
and Furnstahl {\it et al.}\cite{zhang99} (solid).
}
\end{figure}

\begin{figure}
\epsfig{file=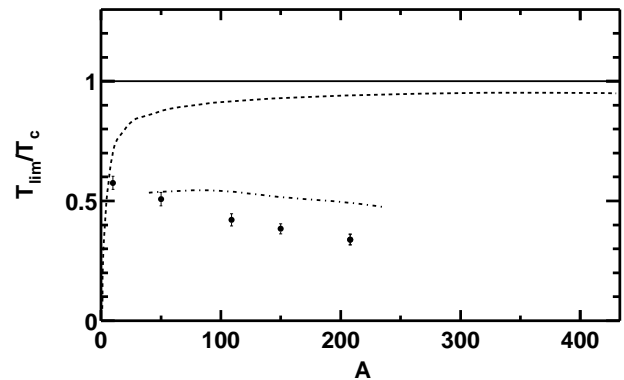,width=9.2cm,angle=0}
\caption{\label{fg:TlimTc}
Theoretical variation of the ratio $T_{lim}/T_c$ with mass along the line 
of beta stability. The solid line indicates the reference value of $T_c$. The 
short dashed line shows the effect of finite size scaling derived from 
an Ising model\cite{elliott00}.  The line with alternating short and long 
dashes depicts the ratio of the nuclear binding energy per nucleon to the 
bulk binding energy per nucleon, 16 MeV. Points with uncertainties are 
derived from the model calculations in references 
\cite{bonche85,levit85,besprovany89,song91,zhang96_1,zhang96_2,%
das92,song93,song94,baldo99,zhang99}.
}
\end{figure}

\begin{figure}
\epsfig{file=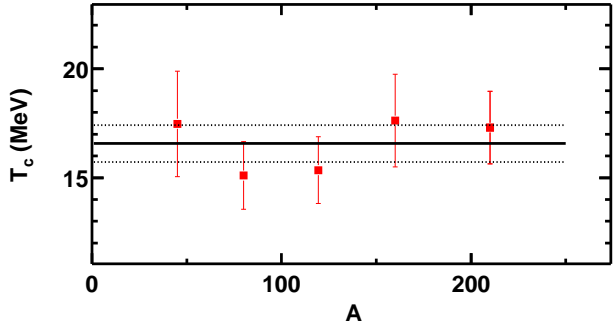,width=9.2cm,angle=0}
\caption{\label{fg:Tc}
Derived values of the critical temperature of symmetric nuclear matter. 
Values derived from data in five different mass regions are presented. 
The mean value of 16.6 MeV is indicated by the horizontal solid line. 
The range corresponding to $\pm$ one standard deviation from this mean value 
is shown by the thin dotted lines.
}
\end{figure}

\begin{figure}
\epsfig{file=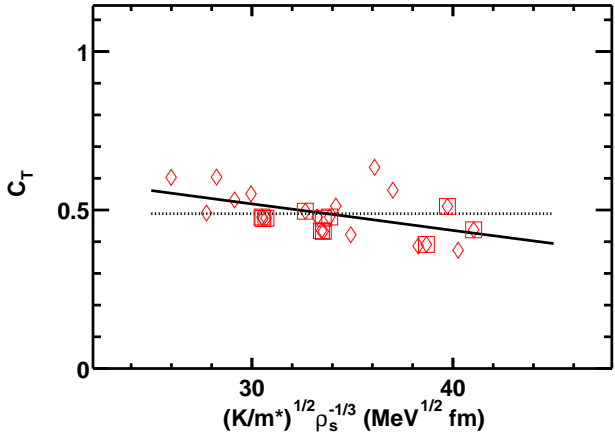,width=9.2cm,angle=0}
\caption{\label{fg:Ct}
The constant $C_T$ of equation (1), evaluated from various microscopic 
calculations.  $C_T$ is plotted against 
$(K/m^*)^\frac{1}{2}\rho_s^{-\frac{1}{3}}$. Derived values of $C_T$ are 
indicated by open diamonds. The dotted horizontal line 
indicates the mean value of $C_T$. The solid line represents the linear 
least squares fit to the derived values. Values of $C_T$ obtained from 
Skyrme and Gogny interactions are further identified by open squares 
placed around the diamonds.
}
\end{figure}

\begin{figure}
\epsfig{file=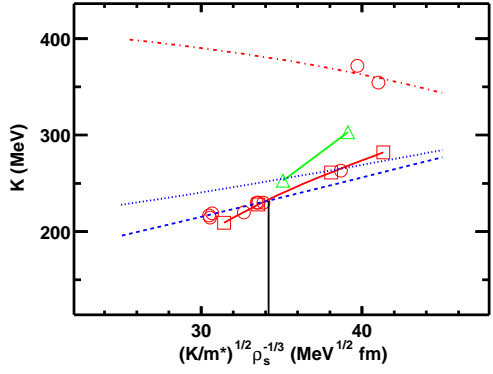,width=9.2cm,angle=0}
\caption{\label{fg:K}
Compressibility modulus, $K$, as a function of 
$(K/m^*)^\frac{1}{2}\rho_s^{-\frac{1}{3}}$. The values obtained for Gogny 
interactions of reference \onlinecite{blaizot95} are represented by 
open squares ($\alpha = \frac{1}{3}$) and open triangles ($\alpha = \frac{2}{3}$). 
Symbols for each set are connected by thin solid lines.  Results using 
different Skyrme interactions are represented by open circles. The other 
lines represent generalized calculations using Skyrme 
interactions \cite{chabanat97} with $\sigma = \frac{1}{6}$(short dashed line), 
$\sigma =\frac{1}{3}$ (dotted line)and $\sigma =1$(alternating dashes and dots).
}
\end{figure}
\end{document}